\documentclass[12pt]{article}
\usepackage{axodraw}
\usepackage{epsfig}
\usepackage{color}
\usepackage{a4}
\usepackage{latexsym}
\usepackage{cite}

\usepackage{pslatex}
\usepackage[latin1]{inputenc}
\usepackage[T1]{fontenc}

\usepackage{color,colordvi}
\usepackage{url}

\def\colour4colour#1{\Blue{#1}}

\newcommand{\gsim}{\raisebox{-0.07cm}{$\:\:\stackrel{>}{{\scriptstyle
 \sim}}\:\: $} }

\newcommand{\equal}{\:\: = \:\:}

\newcommand{\hspn}{{\hspace{-4mm}}}
\newcommand{\hspp}{{\hspace{4mm}}}
\newcommand{\beq}{\begin{equation}}
\newcommand{\eeq}{\end{equation}}
\newcommand{\bea}{\begin{eqnarray}}
\newcommand{\eea}{\end{eqnarray}}
\newcommand{\nn}{\nonumber}
\newcommand{\MSb}{$\overline{\mbox{MS}}$}
\newcommand{\ra}{\rightarrow}

\newcommand{\als}{\alpha_{\rm s}}
\newcommand{\ars}{a_{\rm s}}

\newcommand{\ep}{\varepsilon}

\newcommand{\gone}{{g_{\!1}^{}}}

\begin{document}
\setlength{\parskip}{0.25cm}
\setlength{\baselineskip}{0.54cm}

\def\gamU#1{{\gamma^{\:#1}}}
\def\gamL#1{{\gamma_{\:\!#1}}}
\def\gam5{{\gamma_{\:\!5}^{}}}

\def\bfct#1{{\beta_{\:\! #1}}}
\def\bfcts#1{{\beta_{\:\! #1}^{2}}}

\def\z#1{{\zeta_{#1}}}
\def\zss{\zeta_2^{\,2}}

\def\nc{{n^{}_c}}
\def\ca{{C^{}_A}}
\def\cas{{C^{\,2}_A}}
\def\cat{{C^{\,3}_A}}
\def\cf{{C^{}_F}}
\def\cfs{{C^{\, 2}_F}}
\def\cft{{C^{\, 3}_F}}
\def\nf{{n^{}_{\! f}}}
\def\nfs{{n^{\,2}_{\! f}}}

\def\DNn#1{D_0^{\:#1}}
\def\etaD#1{\eta^{\:\!#1}}

\def\cam2cf{{(\ca-2\cf)}}

\def\as(#1){{\alpha_{\rm s}^{\,#1}}}
\def\ar(#1){{a_{\rm s}^{\,#1}}}

\def\x1{{(1 \! - \! x)}}
\def\LntO{\ln(1\!-\!x)}
\def\Lnt(#1){\ln^{\,#1}(1\!-\!x)}

\def\muRs{{\mu_R^{\,2}}}
\def\Qs{{Q^{\, 2}}}

\def\S(#1){{{S}_{#1}}}
\def\Ss(#1,#2){{{S}_{#1,#2}}}
\def\Sss(#1,#2,#3){{{S}_{#1,#2,#3}}}
\def\Ssss(#1,#2,#3,#4){{{S}_{#1,#2,#3,#4}}}
\def\Sssss(#1,#2,#3,#4,#5){{{S}_{#1,#2,#3,#4,#5}}}

\def\frct#1#2{\mbox{\large{$\frac{#1}{#2}$}}}
\def\frkt#1#2{\mbox{\Large{$\frac{#1}{#2}$}}}

\def\H(#1){{\rm{H}}_{#1}}
\def\Hh(#1,#2){{\rm{H}}_{#1,#2}}
\def\Hhh(#1,#2,#3){{\rm{H}}_{#1,#2,#3}}
\def\Hhhh(#1,#2,#3,#4){{\rm{H}}_{#1,#2,#3,#4}}

\def\Lsxsq{{\ln^{\,2}x}}
\def\Lsx{{\ln x}}
\def\Llx{{\ln \x1}}

\def\dabc2n{{d^{\:\!abc}d_{abc}/\nc}}

\begin{titlepage}
\noindent
% DESY 14--157, \hfill {\tt arXiv:1409.5131}\\
DESY 15-061 \hfill June 2015\\
NIKHEF 15-018 \\ 
LTH 1042 \\
\vspace{1.2cm}
\begin{center}
\Large
{\bf On $\gamma_{\:\!5}$ in higher-order QCD calculations and the } \\
\vspace{0.15cm}
{\bf NNLO evolution of the polarized valence distribution} \\
\vspace{1.6cm}
\large
S. Moch$^{\, a}$, J.A.M. Vermaseren$^{\, b}$ and A. Vogt$^{\, c}$\\
\vspace{1.2cm}
\normalsize
{\it $^a$II.~Institute for Theoretical Physics, Hamburg University\\
\vspace{0.1cm}
D-22761 Hamburg, Germany}\\
\vspace{0.5cm}
{\it $^b$Nikhef Theory Group \\
\vspace{0.1cm}
Science Park 105, 1098 XG Amsterdam, The Netherlands} \\
\vspace{0.5cm}
{\it $^c$Department of Mathematical Sciences, University of Liverpool\\
\vspace{0.1cm}
Liverpool L69 3BX, United Kingdom}\\
\vspace{2.6cm}
\large
{\bf Abstract}
\vspace{-0.2cm}
\end{center}
We discuss the prescription for the Dirac matrix $\gam5$ in dimensional
regularization  used in most second- and third-order QCD calculations of
collider cross sections. We provide an alternative implementation of
this approach that avoids the use of an explicit form of $\gam5$ and of
its (anti-) commutation relations in the most important case of no more
than one $\gam5$ in each fermion trace. 
This treatment is checked by computing the third-order corrections to the 
structure functions $F_2$ and $g_1^{}$ in charged-current deep-inelastic 
scattering with axial-vector couplings to the $W$-bosons. We derive the
so far unknown third-order helicity-difference splitting function $\Delta 
P_{\rm ns}^{\,(2)\rm s}$ that contributes to the next-to-next-to-leading 
order (NNLO) evolution of the polarized valence quark distribution of the
nucleon. This function is negligible at momentum fractions $x\gsim 0.3$
but relevant at $x \ll 1$.
\vfill
\end{titlepage}

% ---------------------------------------------------------------------

\noindent
Dimensional regularization \cite{Bollini:1972ui,'tHooft:1972fi}, i.e., 
the analytic continuation of the theory to a non-integer number $D$ of 
space-time `dimensions' (see also Ref.~\cite{Collins84} for an 
introduction), is the standard framework for higher-order calculations 
in gauge field theories including Quantum Chromo\-dynamics (QCD).
For some semi-leptonic benchmark observables, e.g., in inclusive 
deep-inelastic scattering (DIS) and semi-inclusive $e^+e^-$ annihilation
(SIA), the use of dimensional regularization requires prescriptions for
dealing with the genuinely four-dimensional objects 
 $\ep^{\,(4)}_{\mu\nu\rho\sigma}$,
the totally antisymmetric invariant tensor in four dimensions, and the 
Dirac matrix 
 $\gamma_{\:\!5}^{\,(4)} 
 \,=\, i\, \gamma_0^{} \gamma_1^{} \gamma_2^{} \gamma_3^{}
 \,=\, i/4! \; \ep^{\,(4)}_{\mu\nu\rho\sigma}\,
 \gamU\mu \gamU\nu \gamU\rho \gamU\sigma$.
%
% This is for Willy's metric (F3, 1992), as in Retey/Vermaseren, 
% not that of LarinV (1991). 
 
The tensor $\ep^{}_{\mu\nu\rho\sigma}$ enters in the projection of the 
respective hadronic tensors onto the structure functions $F_{\:\!3}$ 
and $\gone$ in DIS and the fragmentation function $F_A$ in SIA, e.g., 
\beq
\label{WmnA}
  W_{\mu\nu} \equal \ldots \;-\; i\: \ep^{}_{\mu\nu\alpha\beta}\,
  p^{\:\!\alpha} q^{\:\!\beta} \:\frct{1}{pq}\: F_3(x,\Qs) 
\:\: ,
\eeq
%        Willy (1992) = LarinV (1991)
%
where the $x$ is the Bjorken scaling variable, $x = \Qs/(2 pq)$ with
$\Qs = - q^2$, and where we have suppressed all non-$F_{\:\!3}$ parts
of $W_{\mu\nu}$. It also occurs in the helicity-difference projection
of incoming gluons in partonic polarized DIS.
The matrix $\gam5$ enters via the axial-vector coupling of the $W$ and 
$Z$ bosons to the quarks as well as by the corresponding 
helicity-difference projection for quarks. 

In particular the issue of $\gam5$ has attracted a considerable amount of 
attention. The `canonical' approach is that of Ref.~\cite{'tHooft:1972fi}
in which the Dirac algebra, and hence the loop momenta, are split in 
4- and $(D-4)$-dimensional sets with
\bea
\label{HVg5}
  \{ \gam5 , \gamL\mu \} &\!=\!& 0 \;\; , \quad \mu \,=\, 0,\,1,\,2,\,3
\; , \nn \\[0.3mm]
  [ \,\gam5 , \gamL\mu \,]\, &\!=\!& 0 \;\; , \quad \mbox{otherwise}
\:\: ,
\eea
where $\{ a, b \}$ and $[ a , b ]$ denote the standard anti-commutator 
and commutator, respectively.
While Eq.~(\ref{HVg5}) leads to a consistent procedure \cite{g5BM}, it 
has some drawbacks: the occurrence of additional scalar products of 
$(D-4)$-dimensional loop momenta and an intermediate violation of the 
axial Ward identity. 
This situation has triggered quite a few of alternative suggestions which 
we are unable to address in this brief note; the reader is referred to 
\cite{Baikov:1991qz,g5KKS,Trueman:1995ca,Weinzierl:1999xb,%
Jegerlehner:2000dz} and references therein. 

Our focus will be on the scheme developed, on the basis of Ref.~\cite
{Akyeampong:1973xi}, in Refs.~\cite{g5LV,g5L,ZvN-F3} which is closely 
related to that of Refs.~\cite{'tHooft:1972fi,g5BM} but avoids 
complicating the loop integrals. Consequently this scheme has been 
employed in almost all higher-order (next-to-next-to-leading order, 
NNLO, or next-to-next-to-next-to-leading order, N$^3$LO) diagram
calculations of splitting and coefficient functions in DIS \cite
{ZvN-F3,ZvNpol,MochV99,Retey:2000nq,MVV3,MochRogalCC,mvvLL08,MVV10,MVV11} 
and SIA \cite{Rijken:1996npa,Mitov:2006wy}, as well as for the 
determination of the NNLO QCD corrections to the cross section for the 
production of a pseudoscalar Higgs boson 
\cite{Harlander:2002vv,Anastasiou:2002wq,Ravindran:2003um}. 

We have been lead to consider this issue by our work on polarized 
charged-current DIS, in particular the generalization of some of 
Ref.~\cite{SVW95} to the third order, which facilitates the 
determination of the so far unknown NNLO splitting function 
$\Delta P_{\rm ns}^{(2)\rm s}$, the longitudinally polarized analogue of 
$P_{\rm ns}^{(2)\rm s}$ in Ref.~\cite{MVV3}. 
In order to study more cases with more than one $\gam5$ at the three-loop
level, we have redone the  calculations of $F_2$ and $F_L$ of Refs.~\cite
{MochV99,Larin:1994vu,Larin:1997wd} and of $\gone$ of 
Ref.~\cite{mvvLL08} with axial-vector instead of vector couplings to the 
gauge bosons.
In particular for the latter case it was useful to employ an algorithm 
which is equivalent to, but more efficient than, that of 
Refs.~\cite{g5LV,g5L,ZvN-F3}. This alternative implementation 
may be useful for future higher-order calculations in QCD.

% ---------------------------------------------------------------------

In most of the higher-order calculations mentioned above, the 
prescription of Refs.~\cite{g5LV,g5L,ZvN-F3}, sometimes briefly referred 
to as the Larin scheme, has been implemented in the form
\beq
\label{gmu5L}
  (\gamL\mu \gam5)_L^{} \equal \frct{1}{6}\:i\;
  \ep^{}_{\mu\nu\rho\sigma}\, \gamU\nu \gamU\rho \gamU\sigma \; ,
\eeq
i.e., what is continued is the axial-vector matrix, written with a 
specific order of the two matrices. 
Alternatively one can use (as, e.g., in Ref.~\cite{Anastasiou:2002wq})
\beq
\label{g5L}
  \gamma_{\:\!5,L}^{} \equal \frct{1}{4!}\:i\; \ep^{}_{\mu\nu\rho\sigma}
  \,\gamU\mu \gamU\nu \gamU\rho \gamU\sigma \; .
\eeq
Both substitutions, e.g., via Eq.~(\ref{WmnA}), lead to products of 
two $\ep$-tensors which can be evaluated in terms of the $D$-dimensional
metric tensor $\delta_{\,\alpha}^{\;\mu}$ as
\beq
\label{epxep}
  \ep^{\,\mu\nu\rho\sigma}\, \ep^{}_{\alpha\beta\kappa\lambda} \equal
  \left| \begin{array}{cccc}
     \delta_{\,\alpha}^{\;\mu}   & \delta_{\,\beta}^{\;\mu} &
     \delta_{\,\kappa}^{\;\mu}   & \delta_{\,\lambda}^{\;\mu} \\[1mm]
     \delta_{\,\alpha}^{\;\nu}   & \delta_{\,\beta}^{\;\nu} &
     \delta_{\,\kappa}^{\;\nu}   & \delta_{\,\lambda}^{\;\nu} \\[1mm]
     \delta_{\,\alpha}^{\;\rho}  & \delta_{\,\beta}^{\;\rho} &
     \delta_{\,\kappa}^{\;\rho}  & \delta_{\,\lambda}^{\;\rho} \\[1mm]
     \delta_{\,\alpha}^{\;\sigma}& \delta_{\,\beta}^{\;\sigma} &
     \delta_{\,\kappa}^{\;\sigma}& \delta_{\,\lambda}^{\;\sigma} 
   \end{array} \right|
\:\: .
\eeq
%       Anastasiou & Melnikov have a sign here, unlike Willy.
%       It does not really matter ... .
%
The need to use the $D$-dimensional metric on the right-hand side 
has been clearly established, at least for the type of calculations we 
are considering here, in Ref.~\cite{ZvN-F3}.

This implies that the traces 
\beq
\label{Trace1}
  {\rm Tr}\, \left (\gamL{\nu_1} \gamL{\nu_2} \,\ldots\, 
                    \gamL{\nu_{2m-1}} \gamL\mu \, \gam5 \right)
\; ,
\eeq
evaluated using Eqs.~(\ref{gmu5L}) and (\ref{g5L}) are not identical at
$m \geq 3$, as the additional terms generated by Eq.~(\ref{g5L}) cancel
only at $D=4$ due to the Schouten identity, 
\beq
\label{Schouten}
      \ep^{\,(4)}_{\nu_3\nu_4\nu_5\nu_6}\, \delta_{\,\nu_2}^{\:\nu_1}
    + \ep^{\,(4)}_{\nu_4\nu_5\nu_6\nu_2}\, \delta_{\,\nu_3}^{\:\nu_1}
    + \ep^{\,(4)}_{\nu_5\nu_6\nu_2\nu_3}\, \delta_{\,\nu_4}^{\:\nu_1}
    + \ep^{\,(4)}_{\nu_6\nu_2\nu_3\nu_4}\, \delta_{\,\nu_5}^{\:\nu_1}
    + \ep^{\,(4)}_{\nu_2\nu_3\nu_4\nu_5}\, \delta_{\,\nu_6}^{\:\nu_1}
 \equal 0 
\:\: .
\eeq
However, if the above asymmetric non-Hermitian form of the axial-vector 
matrix is replaced by its symmetric Hermitian counterpart (as done  
in Ref.~\cite{Anastasiou:2002wq}),
\beq
\label{Asym}
  \gamL\mu \, \gam5  \:\:\ra\:\:  \frct{1}{2}
  \left( \gamL\mu \, \gam5 \,-\, \gam5 \, \gamL\mu \right)
\:\: ,
\eeq
then Eq.~(\ref{g5L}) leads to exactly the same results as 
Eq.~(\ref{gmu5L}) for the trace (\ref{Trace1}). 
The situation is completely analogous if the prescriptions
(\ref{gmu5L}) and (\ref{g5L}) are applied, for any $m$ and $n$, to 
\beq
\label{Trace2}
  {\rm Tr}\, \left (\gamL{\nu_1} \gamL{\nu_2} \,\ldots\,
                    \gamL{\nu_{m}} \gamL{\mu_1} \gam5 \, 
                    \gamL{\rho_1} \gamL{\rho_2} \,\ldots\,
                    \gamL{\rho_{n}} \gamL{\mu_{\:\!2}} \gam5 \right)
\:\: :
\eeq
Eq.~(\ref{gmu5L}) and Eq.~(\ref{g5L}) with (\ref{Asym}) lead to the 
same results. Cases with more $\gam5$ will be addressed below.

Obviously the inconsistent use of Eq.~(\ref{g5L}) without 
Eq.~(\ref{Asym}) leads to wrong results in diagram calculations only 
in sufficiently complicated cases. For example, re-calculating the 
third-order corrections for $F_3^{}$ \cite{Retey:2000nq,MVV10} in this
manner leads to the same results as Eq.~(\ref{gmu5L}) for each 
individual diagram including its dependence on the gauge parameter.
On the other hand, wrong (and unfactorizable, cf.~Ref.~\cite{ZvN-F3}) 
results would be obtained for the polarized vector--axialvector 
interference structure functions $g_{4,5}^{}$ (using the labeling 
conventions of Ref.~\cite{Agashe:2014kda}) in which $\gam5$ occurs 
not once, as for $F_3$, but twice.

While the calculation of the Dirac traces is not usually a limiting
factor in higher-order calculations, the introduction of additional
matrices by Eq.~(\ref{gmu5L}) or  Eq.~(\ref{g5L}) has sometimes been
considered a drawback of the Larin scheme. For traces with one $\gam5$, 
the most important case in QCD calculations (e.g., the only one 
encountered in Refs.~\cite{ZvN-F3,ZvNpol,MochV99,Retey:2000nq,MVV3,%
MochRogalCC,mvvLL08,MVV10,MVV11,Rijken:1996npa,Mitov:2006wy,%
Harlander:2002vv,Anastasiou:2002wq,Ravindran:2003um}),
this issue can be avoided by using algorithms which are completely
equivalent and do not introduce any additional intermediate matrices.

\noindent
A procedure equivalent to, but faster than, using Eq.~(\ref{gmu5L}) 
is provided by the following steps:
\vspace*{-2mm}
\begin{enumerate}
\item Write the one-$\gam5$ traces in the form (\ref{Trace1}) without 
 changing the order of the $\gamma$-matrices. This can be viewed as 
 using the cyclicity of the trace, or as reading it from this point, 
 cf.~Ref.~\cite{g5KKS}.
\item Evaluate Eq.~(\ref{Trace1}) using
\bea
 {\rm Tr}\:  \left (\gamL{\nu_1} \gamL{\nu_2} \,\ldots\,
             \gamL{\nu_{2m-1}} \gamL\mu \, \gam5 \right) 
 &\!=\!& \mbox{} -4\,i\:
   g_{\nu_1 \nu_2}^{} \ldots g_{\nu_{2m-5}\nu_{2m-4}}^{}
   \ep_{\:\!\nu_{2m-3} \nu_{2m-2} \nu_{2m-1} \mu}^{}
\nn  \\[0.5mm] & & \mbox{}
   \pm \mbox{ permutations of~ } \nu_1^{} \ldots \nu_{2m-1}^{}
\:\: .
\label{gmu5Limpl}
\eea
Incidentally, this main step can be programmed in {\sc Form} 
\cite{FORM3,TFORM,FORM4}, for an extensive documentation see 
\cite{FORMdoc}, in a very compact manner for any number of traces 
with one $\gam5$, viz
\begin{verbatim}
repeat;
   id,once,G(m1?,?a,mu?,five) = distrib_(-2,3,G1,G2,?a)*G(mu,five);
   id G2(mu1?,mu2?,mu3?)*G(mu4?,five) = e_(mu1,...,mu4);
endrepeat;
.sort
repeat;
   if ( count(G1,1) );
       id,once,G1(?a) = g_(1,?a);
       Tracen,1;
   endif;
endrepeat;
\end{verbatim} 
\item For traces with more than one $\gam5$, use Eq.~(\ref{gmu5L}) 
 for all but one (special care is needed for more than two $\gam5$, 
 see below), then calculate the resulting  one-$\gam5$ trace according 
 to 1.~and 2.~above. 
\end{enumerate}

\noindent
A corresponding algorithm equivalent to Eq.~(\ref{g5L}) can be 
implemented by changing 1.~and 2.~above~to
\vspace*{-6mm}
\begin{enumerate}
\item Input all axial-vector matrices in the form (\ref{Asym}), then
 proceed as under 1.~above. 
\item Evaluate the resulting traces, in which now $\gamL\mu$ has no 
special role, as
\bea
 {\rm Tr}\:  \left (\gamL{\nu_1} \gamL{\nu_2} \,\ldots\,
             \gamL{\nu_{2m-1}} \gamL\mu \, \gam5 \right)
 &\!=\!& \mbox{} -4\,i\:
   g_{\nu_1 \nu_2}^{} \ldots g_{\nu_{2m-5}\nu_{2m-4}}^{}
   \ep_{\:\!\nu_{2m-3} \nu_{2m-2} \nu_{2m-1} \mu}^{}
\nn  \\[0.5mm] & & \mbox{}
   \pm \mbox{ permutations of~ } 
   \nu_1^{} \ldots \nu_{2m-1}^{}\, \mu
\:\: ,
\label{g5Limpl}
\eea
for which the central two lines of the above {\sc Form} implementation 
are changed to the simpler
\begin{verbatim}
    id,once,G(m1?,?a,five) = distrib_(-2,4,G1,G2,?a);
    id  G2(mu1?,...,mu4?) = e_(mu1,...,mu4);
\end{verbatim}
\end{enumerate}
%\pagebreak
%
Eq.~(\ref{g5Limpl}) has certainly been used elsewhere before; however 
we have not seen a clear discussion of the `implicit-$\gam5$' relations 
(\ref{gmu5Limpl}) and (\ref{g5Limpl}) to the `explicit' prescriptions 
(\ref{gmu5L}) and (\ref{g5L}) in the literature.
 
We now have four equivalent manners to evaluate traces with $\gam5$
and should briefly address their efficiency: 
computing Eq.~(\ref{Trace1}) for $m=7$, i.e., with 14 $\gamma$-matrices
besides $\gam5$, requires about 1.2 and 38 seconds, respectively, using
(\ref{gmu5L}) and (\ref{g5L}) with the internal trace algorithms of
{\sc Form}, but 0.3 and 1.2 seconds with the shown implementations of 
Eqs.~(\ref{gmu5Limpl}) and (\ref{g5Limpl}) on a Xeon E5-2667v2 with 
3.30GHz, using one core. 
The corresponding numbers for $m=8$ are higher by factors of about~20.
This scaling is the same as for the non-$\gam5$ case, which is however
faster by almost a factor of 8 than our fastest $\gam5$ implementation
(\ref{gmu5Limpl}). The corresponding execution times for 
Eq.~(\ref{Trace2}) with $m=n=5$ (12 $\gamma$-matrices besides the two
$\gam5$) are, in the same order, 4.5, 740, 1.3 and 55 seconds; the two 
faster methods again take longer by about a factor of 20 for $m=n=6$.

We now move to the application of the above $\gam5$ scheme in 
higher-order calculations, focusing on the best known (in general and 
to us) case of third-order DIS in massless perturbative QCD. 
This scheme shares the second drawback of the `t Hooft$/$Veltman
scheme (\ref{HVg5}), the violation of the axial Ward identity.
This issue is less serious here than it may be in higher-order 
calculations in the electroweak theory; it is addressed by 
`correcting' the axial current by the renormalization factors $Z_5$
and $Z_A$ determined to the third order in the strong coupling constant
$\als$ in Ref.~\cite{g5LV}, 
\bea
\label{ZA}
  Z_{A} &\!=\!& 
%%START
%%L %%texZA =
  1
  + \ar(2) \, \* 
    \ep^{-1} \: \* 2\,\* \cf \* \bfct0 
  - \ar(3) \* \left[ 
                     \ep^{-2} \: \* \frct{4}{3}\: \* \cf \* \bfcts0
                   - \ep^{-1} \: \* \frct{2}{9}\: \* \cf \* 
%%STOP
    \!
%%START
    \left( 6\,\* \bfct1 + \bfcts0 - 42\,\* \cf\*\bfct0 
        + 32\,\* \ca\*\bfct0 \right) 
%%STOP
    \! 
%%START
    \right]
%%;
%%STOP
 \, , \quad \\[1mm] 
\label{Z5}
  Z_{5} &\!=\!& 
%%START
%%L %%texZ5 =
  1
  - \ars\: \* 4\, \* \cf 
  + \ar(2) \* \bigg[ 
            22\, \* \cfs
          - \frct{107}{9}\: \* \cf \* \ca
          + \frct{2}{9}\: \* \cf \* \nf
  \bigg]
%%STOP
\nonumber \\[0.5mm] & & \mbox{} 
%%START
  + \ar(3) \* \bigg[
            \cft \* \bigg(
          - \frct{370}{3} 
          + 96\, \* \z3 \bigg)
          + \cfs \* \ca \* \bigg( \:
            \frct{5834}{27} 
          - 160\, \* \z3 \bigg)
          + \cf \* \cas \* \bigg(
          - \frct{2147}{27} 
          + 56\, \* \z3 \bigg)
%%STOP
\nonumber \\[0.5mm] & & \mbox{} 
%%START
          + \cfs \* \nf \* \bigg(
          - \frct{62}{27} 
          - \frct{32}{3}\: \* \z3 \bigg)
          + \ca \* \cf \* \nf \* \bigg( \:
            \frct{356}{81} 
          + \frct{32}{3}\: \* \z3 \bigg)
          + \frkt{52}{81}\: \* \cf \* \nfs \,
  \bigg]
%%;
%%STOP
\end{eqnarray}
for $D = 4 - 2\:\!\ep$.
These factors are expressed in terms of the renormalized coupling
normalized as $\ars = \als/(4\pi)$, and we have employed the first 
two coefficients of the $\beta$-function of QCD
\cite{beta0a,beta0b,beta1a,beta1b},
\beq
  \bfct0 \equal \frct{11}{3}\: \ca - \frct{2}{3}\: \nf
\;\: , \quad
  \bfct1 \equal \frct{34}{3}\: \cas - \frct{10}{3}\: \ca \nf
  -  \frct{2}{3}\: \cf\nf
\;\: .
\eeq
to write Eq.~(\ref{ZA}) in a slightly more compact form. 

On top of, or instead of, the multiplication with $Z_5\,Z_A$ before 
performing the mass factorization, a non-trivial factorization-scheme 
transformation is required in the polarized case for arriving at the 
splitting and coefficient functions in \MSb\ for the helicity-dependent 
case, see Refs.~\cite{MvN95,WVdP1a,WVdP1b,MSvN98,mvvLL08,MVV11}.
At N$^3$LO this transformation is not fully known yet: the 
pure-singlet quark contribution is missing.

A second yet innocuous effect of using Eqs.~(\ref{gmu5L}) -- 
(\ref{epxep}) (or any equivalent algorithm) is that all traces, 
including those of the $\as(0)$ Born contributions, receive an 
additional dependence on $D$.  
This dependence is factorized and then removed in the projection on
the structure functions. E.g., the well-known $D$-dependence in the 
projection on the structure function $F_3$,
\beq
\label{projF3}
 P_{3}^{\,\mu\nu} \equal - {\rm{i}}\: \frkt{1}{(D-3)(D-2)}
 \; \ep^{\,\mu\nu\alpha\beta}\: \frac{p_\alpha q_\beta}{p\cdot q}
\:\: ,
\eeq
originates in the basic trace of $\gam5$ with four other $\gamma$-%
matrices and $\ep_{\mu\nu pq}^{}\, \ep^{\,\mu\nu pq} \sim (D-2)(D-3)$.
This factor is analogous to the $(D-2)^{-1}$ in the $F_2$ projection 
that arises from $\gamL\rho \gamL\mu\,\gamU\rho \,=\, (2-D)\,\gamL\mu$.

\begin{figure}[htb]
\vspace{2mm}
\centerline{\epsfig{file=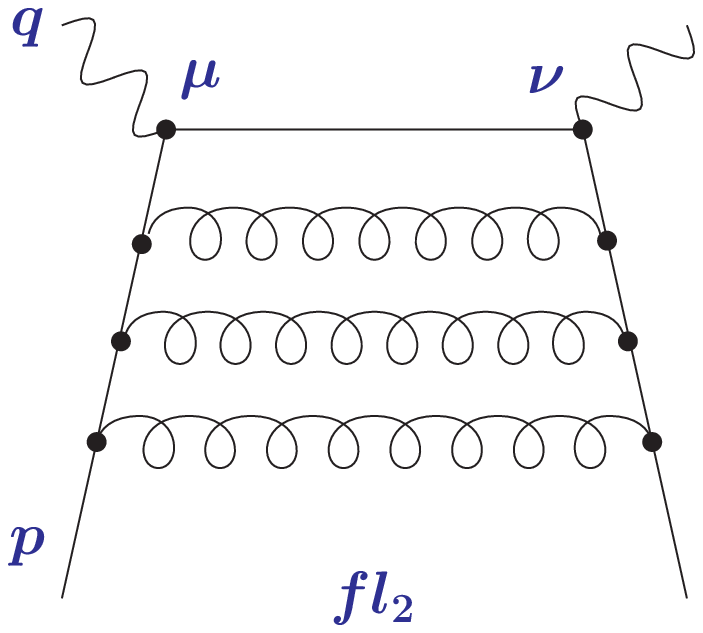,width=6cm}\hspace*{12mm}
\epsfig{file=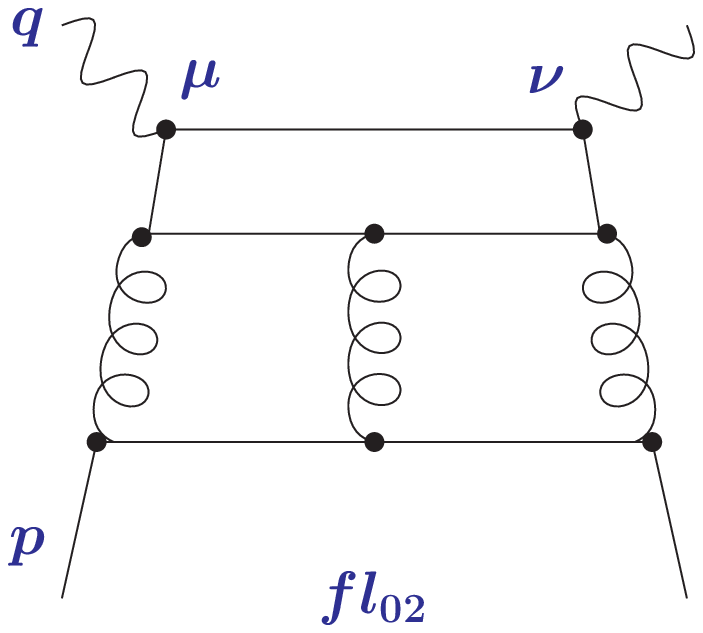,width=6cm}}
\vspace*{-1mm}
\caption[]{\label{fig:3loopDIS}
Typical third-order Feynman diagrams for the two flavour classes
contributing to quark-initiated charged-current DIS. Depending on
the structure function, the boson lines at the top are replaced by
$\ep_{p q \mu\nu}$ or a combination of $g_{\mu\nu}^{}$ and
$p_\mu p_\nu^{}$, and the quark lines at the bottom by $\gamL{p}$
or $\gamL{p} \gam5$ (in Schoonship notation).
The vertices with $\mu$ and $\nu$ represent vector or axial-vector
couplings.
}
\vspace*{1mm}
\end{figure}

As mentioned above, Eq.~(\ref{gmu5L}) has been extensively used in 
higher-order QCD correction in cases where only one $\gam5$ occurs.
On the other hand, we are not aware of a corresponding NNLO or N$^3$LO 
calculation involving two occurrences $\gam5$ in either the same or 
different traces.
The former case is more interesting and challenging; a good first
example is a third-order calculation of the structure functions 
$F_2$ and $F_L$, for typical forward-Compton diagrams see 
Fig.~\ref{fig:3loopDIS}, with an axial-vector instead of the vector 
coupling \cite{Larin:1994vu,Larin:1997wd,MVV4,MVV6} to the gauge boson.

Since we are not looking for a new splitting or coefficient function
in this calculation, it is sufficient to keep the full dependence on
the Mellin variable $N$ only to two loops and determine the third-order
corrections for a few even-integer values of $N$. At this level the
computation is straightforward and virtually automatic; we can use 
our old diagrams databases and employ our calculation and analysis 
programs with minor modifications.

The projection on the axial-vector structure functions
$F_2$ and $F_L$ involves a different prefactor. The Born-level trace
($p/q$ are the quark$/$gauge-boson momenta, $p^{\,2}=0$) is now
\beq
  {\rm Tr} \left (\gamL{p}\, \gamL\mu \gam5\, \gamL{p+q}\, 
  \gamU\mu\:\! \gam5 \right) ~\sim~ (D-1) (D-2) (D-3) (D-6)\, pq
\;\; ,
\eeq
%
% Prefactor: 2/3
%
i.e., the projections involve an extra factor $[(D-1)(D-3)(6-D)]^{-1}$.
Taking this into account, and multiplying the results by $(Z_5\, Z_A)^2$
as given by Eq.~(\ref{ZA}) and (\ref{Z5}) before factorization, we 
obtain the same splitting functions and quark and gluon 
coefficient functions for $F_2$ and $F_L$ as found before,
\beq
\label{c2Laa}
  c_{i,\rm q/g}^{\,(n)\rm aa}(x) \equal c_{i,\rm q/g}^{\,(n)\rm vv}(x)
\quad \mbox{ for } \quad  i \:=\: 2,\,L 
\quad \mbox{ at }  \quad  n \:=\: 1,\,2,\,3 
\:\: .
\eeq
This demonstrates that there is no need to resort, as often done, to 
a fully anti-commuting $\gam5$ in traces with two $\gam5$ (which 
admittedly would lead to the right result here): 
the scheme considered here  can be used also for these cases, at a 
usually tolerable cost in computing~time.

The ultimate $\gam5$ challenge, in the framework of QCD corrections for
structure functions at the lowest order in the electroweak theory, is
provided by doing the same for structure function $\gone$ in polarized 
DIS. This calculation involves, in addition to that for $F_2$, an 
$\ep$-tensor from the projection on $\gone$, essentially the same as 
Eq.~(\ref{projF3}), and a $\gam5$ or $\ep$-tensor taking the quark or 
gluon helicity difference.
The resulting contractions of four $\ep$-tensors have to be performed
with special care as their order matters in the present case, unlike in
four dimensions where the results can be shown to be the same by 
repeated application of the Schouten identity (\ref{Schouten}), 
cf.~Ref.~\cite{FORMep4}.
 
In the case at hand, it is correct to pair the $\ep$-tensors from the 
axial-vector vertices 
(labelled $\mu$ and $\nu$ in Fig.~\ref{fig:3loopDIS}).
This is readily achieved in {\sc Form} by using the built-in tensor
{\tt e}$_{-}$ only for these, and to `protect' the other two $\ep$-%
tensors by using a different notation until the other contractions 
and traces have been performed. The fastest implementation is to use
Eq.~(\ref{gmu5L}) for the axial-vector gauge-boson vertices together 
with Eq.~(\ref{gmu5Limpl}), with {\tt e}$_{-}$ suitably renamed in the 
{\sc Form} code shown below that equation. A three-fold application of
Eq.~(\ref{gmu5L}) is also possible, if considerably slower. 
The symmetric implementations (\ref{g5L}) and (\ref{g5Limpl}) are yet 
less efficient; the four-fold application of (\ref{g5L}) is 
prohibitively slow at the third order. 
Only now all four prescriptions consistently lead to
\beq
  {\rm Tr}  \left( \gamL{p}\gam5\, \gamL\mu \gam5\, \gamL{p+q}\,
  \gamL\nu\:\! \gam5 \right)\, \ep^{p q \mu\nu}
  ~\sim~ (D-2) (D-3)^2 (D-6)\, (pq)^2
 \:\: ,
\eeq
%
% Prefactor: 2
%
enabling us to verify, by diagram calculations,
\beq
\label{cg1aa}
  c_{g_1,\rm q/g}^{\,(n)\rm aa}(x) \equal c_{g_1,\rm q/g}^{\,(n)\rm vv}(x)
\quad \mbox{ at }  \quad  n \:=\: 1,\,2,\,3
\:\: .
\eeq

Finally we address the splitting functions $\Delta P_{\rm ns}^{\,-,\rm v}$
for the polarized quark-antiquark differences 
\bea
\label{Dqmin}
  \Delta f_{ik}^{\,-} &\! = \!&
  \Delta f_{q_i^{}} - \Delta f_{\bar{q}_i^{}}
  - \left( \Delta f_{q_k^{}} - \Delta f_{\bar{q}_k^{}} \right)
\:\: , \\
  \Delta f^{\,\rm v} &\! = \!&
  \sum_{i=1}^{\nf} \left\{ \Delta f_{q_i^{}}
                         - \Delta f_{{\bar q}_i^{}} \right\}
\label{Dqval}
\eea
of helicity-dependent parton distributions,
$ \Delta f_{i} \,=\, f_{i}^{\, +} - f_{i}^{\,-} $, 
where $f_i^{\,+}$ and $f_i^{\,-}$ represent the distributions of the 
parton $i$ with positive and negative helicity, respectively, in a 
nucleon with positive helicity, and $\nf$ is the number of 
effectively massless flavours.
For general reasons one expects also at NNLO, $n=2$, a direct relation
between the polarized and unpolarized non-singlet cases
\beq
\label{DPns-}
   \Delta P_{\rm ns}^{\,-(n)} \equal P_{\rm ns}^{\,+(n)} \;\; , 
\eeq
of which the right-hand side was calculated to NNLO in Ref.~\cite{MVV3}.
On the other hand, the difference 
\beq
\label{DPnsS}
   \Delta P_{\rm ns}^{\,\rm s}  \equal
   \Delta P_{\rm ns}^{\,\rm v} - \Delta P_{\rm ns}^{\,\rm -}
\eeq
can only be determined by a diagram calculation. It is this calculation,
via the two-$\gam5$ polarized vector-axialvector interference structure 
function $g_5^{}$ (cf. Ref.~\cite{Agashe:2014kda}) that lead to our above considerations on $\gam5$. 
In particular, $\Delta P_{\rm ns}^{\,(2)\rm s}$ is obtained from the 
flavour class  $fl_{02}$ in Fig.~\ref{fig:3loopDIS}, where the $W$-bosons 
are not attached to the external quark line, for the helicity projection 
$p^{\,\mu}\gamL\mu \gam5 \,\equiv\, \gamL{p} \gam5$ and the structure 
function projection $g_{\mu\nu}^{}$, i.e., with the two $\gam5$ entering 
in different traces. 

The resulting even-$N$ Mellin-space expression reads
\bea
\label{DPnsS2N}
  && \hspn\hspn\hspn \Delta P_{\rm ns}^{\,(2)\rm s}(N) \equal
%%START
%%L %%texDPns2SN = 
  16\, \* \nf\, \* \dabc2n \*  \Big(
         \S(-3)\,  \* (
          - 20\,\*\eta
          + 8\,\*\etaD2
          )
       + \Ss(1,-2)\,  \* (
            8\,\*\eta
          - 16\,\*\etaD2
          )
%%STOP
\qquad \nn\\[-0.5mm] && \mbox{}
%%START
       + 32\,\*\eta\, \* \Ss(-2,1)
       + \S(3)\,  \* (
            6\,\*\eta
          + 4\,\*\etaD2
          )
       + \S(-2)\,  \* (
            8\,\*\eta
          + 20\,\*\etaD2
          + 8\,\*\etaD3
          - 4\,\*\DNn2
          )
%%STOP
\qquad \nn\\[0.5mm] && \mbox{}
%%START
       + \S(1)\,  \* (
            8\,\*\eta
          - 14\,\*\etaD2
          - 42\,\*\etaD3
          - 12\,\*\etaD4
          - 2\,\*\DNn2
          )
       + \DNn2
       \Big)
%%;
%%STOP
\eea
in the notation of Ref.~\cite{MVV11}, i.e., with $D_k = (N+k)^{-1}$, 
$\eta = D_0^{} D_1^{}$ and all harmonic sums \cite{Hsums} taken at 
argument $N$.  The corresponding $x$-space result, in terms of harmonic 
polylogarithms \cite{HPLs} at argument $x$ (also suppressed), is given by
\bea
\label{DPnsS2x}
  && \hspn \Delta P_{\rm ns}^{\,(2)\rm s}(x) \equal
%%START
%%L %%texDPns2Sx = 
  16\, \* \nf\, \* \dabc2n \* \Big(   
       (1-x)\* (
         (24 - 20\,\*\z2) \* \H(1)
       - 8 \* \z2\, \* \Hh(0,-1)
       - 2\, \* \Hhh(1,0,0)
%%STOP
\qquad \nn\\ && \mbox{}
%%START
       - 16\, \* \Hhhh(0,-1,-1,0)
       + 8\, \* \Hhhh(0,-1,0,0)
       + 8\, \* \Hhhh(0,0,-1,0)
       )
     + (1+x)\* (
         16\, \* \Hh(-1,0)
       - 52 \* \z2\, \* \H(-1)
%%STOP
\qquad \nn\\[0.5mm] && \mbox{}
%%START
       - 8 \* \z2\, \* \Hh(0,1)
       - 40\, \* \Hhh(-1,-1,0)
       + 36\, \* \Hhh(-1,0,0)
       + 32\, \* \Hhh(-1,0,1)
       + 12\, \* \Hhhh(0,0,0,1)
       - 4\, \* \Hhhh(0,1,0,0)
       )
%%STOP
\qquad \nn\\[0.5mm] && \mbox{}
%%START
     - x\,\* (
         16\, \* \z3\, \* \H(0)
       + 8\, \* \Hh(0,0)
       + 36\, \* \Hhh(0,0,0)
       - 8\, \* \Hhhh(0,0,0,0)
       )
       - \H(0) \* (
            1
          + 24\,\*x
          )
       - \H(0) \* (
            6
          - 74\,\*x
          )\*\z2
%%STOP
\qquad \nn\\[0.5mm] && \mbox{}
%%START
       - \Hh(0,0) \* (
            12
          + 20\,\*x
          )\*\z2
       + \Hh(0,1) \* (
            10
          + 8\,\*x
          )
       + \Hhh(0,-1,0) \* (
            8
          + 36\,\*x
          )
       + \Hhh(0,0,1) \* (
            6
          - 38\,\*x
          )
%%STOP
\qquad \nn\\ && \mbox{}
%%START
      - (   10
          - 8\,\*x
          ) \* \z2
       + (  6
          + 88\,\*x
          ) \* \z3
       + (  25
          + 15\,\*x
          ) \* \z4
     \Big)  
%%;
%%STOP
\:\: ,
\eea
which can be parametrized, with an accuracy of about 0.1\% or better 
for $10^{\,-6} \leq x \leq 0.95$, by
\bea
\label{dPps2Spar}
  \Delta P_{\rm ns}^{\,(2)\rm s}(x) &\!\cong\!&
  \nf \, (1-x) \left(
   - 42.97\,\* L_0^2 - 29.29\,\* L_0 
   + 179.1 + 117.8\,\* x - 385.5\,\* x^2 
   + 75.94\,\* x^3
\right. \nn \\ & & \hspp \left. \mbox{}
   +  x\,\*L_0\, \*  (8.818\,\* L_0 + 460.8) + 2.681\,\* \Llx
  \right)
   + 0.0001 \,\nf \, \delta\x1
\;\; , \quad
\eea
%      dP2S:  - 42.97* L*L - 29.29* L + 179.1 + 117.8* x - 385.5* x*x
%    ,        + 75.94* x**3 + 460.8* x*L + 8.818* x*L*L + 2.681* L1
%
where $L_0=\ln x$ and 
the artificial $\delta\x1$ contribution can be included to 
compensate the slightly lesser accuracy at very large $x$ to improve 
the approximation for high-$N$ moments and large-$x$ convolutions.
Eqs.~(\ref{DPnsS2N}) and (\ref{DPnsS2x}), together with our calculational
verification of Eq.~(\ref{DPns-}) from the even moments of $\gone$ 
at NNLO, cf.\ Ref.~\cite{SVW95}, complete the determination of the 
third-order helicity-dependent splitting functions of which the main part
was performed in Ref.~\cite{MVV11}.

Eq.~(\ref{DPnsS2x}) can be employed to determine also the odd moments,
in particular
\beq
\label{DPnsS2N1}
  \Delta P_{\rm ns}^{\,(2)\rm v}(N=1) \equal 
%%START
%%L %%texDPns2SN1 =
  8\* \nf\, \* \dabc2n \*
  \,\left( \, 23 \:-\: 12\,\*\z2 \:-\: 16\,\*\z3 \right)
%%;
%%STOP
\:\: .
\eeq
Together with
\bea
\label{DPnsMN1}
  \Delta P_{\rm ns}^{\,(1)\rm -}(N\!=\!1) &\!=\!&
%%START
%%L %%texDPns1mN1 =
  \cf\, \* (\ca - 2\*\cf)\, \* 
    \left( - 13 + 12\,\* \z2 - 8\,\*\z3 \right)
%%;
%%STOP
\:\: , \\[1mm]
  \Delta P_{\rm ns}^{\,(2)\rm -}(N\!=\!1) &\!=\!&
%%START
%%L %%texDPns2mN1 =
    \cfs\, \* \cam2cf\, \* \Big( \,
           \frkt{145}{2} 
         - 62 \* \z2 
         + 164\,\* \z3
         - 372\,\* \z4
         + 48\,\* \z2\*\z3
         + 208\,\* \z5
         \Big)
%%STOP
\nn \\[1mm] & & \hspn\hspn \mbox{}
%%START
  + \cf\*\ca\, \* \cam2cf \* \Big( \,
           \frkt{1081}{36} 
         + \frkt{245}{3}\,\*\z2
         - \frkt{3214}{9}\,\*\z3
         + \frkt{1058}{3}\,\*\z4
         - 48\,\*\z2\*\z3
         - 112 \*\z5
         \Big)
%%STOP
\nn \\[1mm] & & \hspn\hspn \mbox{}
%%START
  - \,\cf\*\nf\, \* \cam2cf\, \* \Big( \,
           \frkt{76}{9} 
         + \frkt{44}{3}\,\* \z2 
         - \frkt{448}{9}\,\* \z3
         + \frkt{68}{3}\,\* \z4
         \Big)
%%;
%%STOP
\label{DPnsMN2}
\eea
-- note the presence of $\z2$ and the higher weight in the Riemann-$\zeta$
function as compared to the `natural' even moments, cf.~Section 3 of 
Ref.~\cite{MRV2007} -- this leads to the expansion
\beq
\label{numexp}
  \Delta P_{\rm ns}^{\,(2)\rm v}(N\!=\!1) ~\cong~ \mbox{}
    -\, 0.00810\, \als^2 
  \,-\, \left( 0.04075 - 0.01850\,\nf \right) \als^3 
  \,+\, {\cal O}(\als^4)
\eeq
in QCD, i.e., for $\,\ca = \nc = 3$, $\,\cf = 4/3\,$ and 
$d^{abc}d_{abc}/\nc = 5/18$.
For the normalization of the latter we had to choose between the
earlier convention of Refs.~\cite{g5LV,Larin:1997wd,Retey:2000nq,MVV3} 
and that of Refs.~\cite{MVV6,MVV10} and other more recent articles 
following Eq.~(187) of Ref.~\cite{vanRitbergen:1998pn}. 
We have done the latter; however the reader should be aware that the 
unpolarized counterparts of Eqs.~(\ref{DPnsS2N}) -- (\ref{DPnsS2x}) 
were presented using the convention $d^{abc}d_{abc}/\nc = 40/9$ in 
Ref.~\cite{MVV3}.

Returning to Eq.~(\ref{numexp}), we note that for $\nf = 4$ 
Eq.~(\ref{DPnsS2N1}) provides almost two thirds of the $\als^3$ 
correction, which is actually larger than the tiny $\als^2$ part at 
normal scales; without it the coefficient of $\als^3\, \nf$ would only 
amount to 0.00061. 
At large-$x$ $\Delta P_{\rm ns}^{\,\rm s}$ is negligible though:
it is suppressed by two powers of $\x1$ with respect to $\Delta 
P_{\rm ns}^{\,\rm \pm}$, with the leading large-$x$ term the same 
as for its unpolarized counterpart in Eq.~(4.11) of Ref.~\cite{MVV3} 
with $\, 
%%START
%%L %%texDPns2Sx1 =
32\,\* \nf\,\* \dabc2n \* (2\*\z2 - 3 ) \* \x1 \* \Llx
%%;
%%STOP
$.

\begin{figure}[htb]
\vspace*{-2mm}
\centerline{\hspace*{-1mm}\epsfig{file=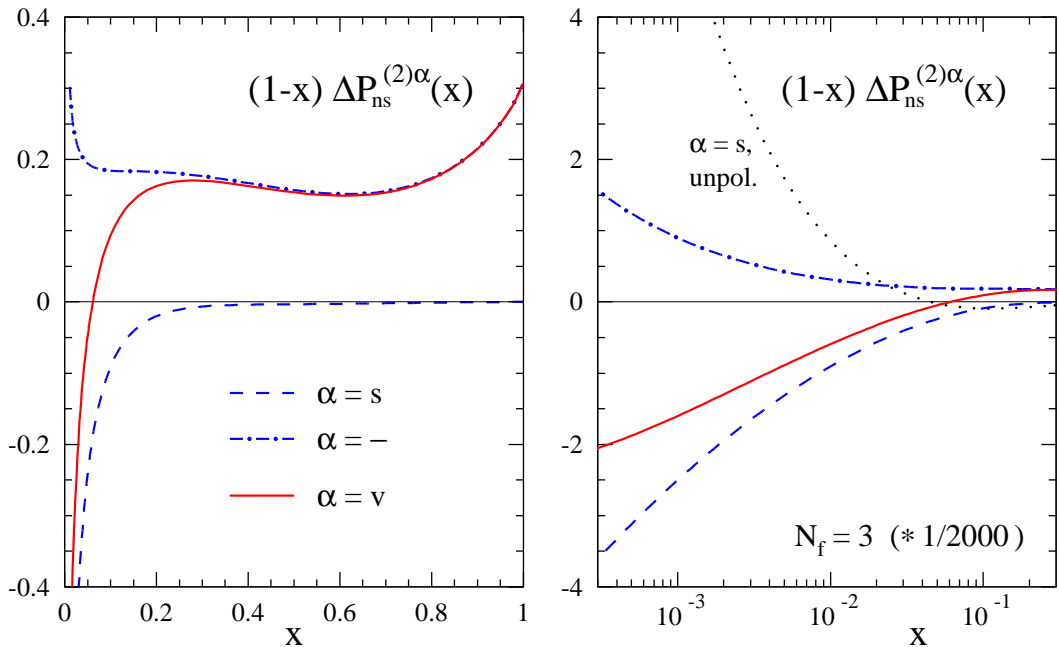,width=14.5cm}}
\vspace{-2mm}
\caption{\label{dPnsMVFig}
The NNLO splitting functions $\Delta P_{\rm ns}^{\,(2)-}$ and
$\Delta P_{\rm ns}^{\,(2)\rm v}$ for the polarized quark distributions
(\ref{Dqmin}) and (\ref{Dqval}), together with the previously unknown 
leading contribution (\ref{DPnsS2x}) to their difference (\ref{DPnsS})
for three flavours, divided by $2000 \simeq (4\,\pi)^3$  to compensate 
for our small expansion parameter $\ars = \als/(4\,\pi)$.
Also shown, on the right, is the unpolarized counterpart 
$P_{\rm ns}^{\,(2)\rm s}(x)$ \cite{MVV3} of Eq.~(\ref{DPnsS2x}).
}
\end{figure}

The situation is totally different at small $x$, as shown in
Fig.~\ref{dPnsMVFig}: despite an only quadratically logarithmic
(negative) small-$x$ enhancement,
\bea
  \Delta P_{\rm ns}^{\,(2)\rm s}(x) &\!=\!&
%%START
%%L %%texDPns2Sx0 =
  - 16\,\* \nf\, \* \dabc2n \* \left\{ \, 6\,\* \z2\, \* L_0^2
  \:+\: ( 1 + 6\,\*\z2 ) \* L_0
%%STOP
  \:+\: {\cal O}(1) 
%%START
  \right\}
%%;
%%STOP
\:\: ,
\eea
its coefficients are such that it overwhelms at $x > 10^{\,-6}$ the
(positive) small-$x$ behaviour of $\Delta P_{\rm ns}^{\rm -}$ which
includes terms up to $\ln^{\,4} x$ that are, due to Eq.~(\ref{DPns-}),
given by Eq.~(4.15) of Ref.~\cite{MVV3}.

To summarize, we have discussed some subtleties of the (multiple) use 
of the $\gam5$ prescription of Refs.~\cite{g5LV,g5L} with a 
$D$-dimensionally contracted $\ep$-tensor \cite{ZvN-F3} in higher-order 
QCD calculations, and provided a procedure that is considerably faster 
than the algorithm mostly used so far and hence may be useful in some 
future three- and four-loop calculations in QCD. 
We have applied our findings to re-derive some third-order results in 
polarized and unpolarized deep-inelastic scattering, and to calculate 
the hitherto unknown NNLO splitting function 
$\Delta P_{\rm ns}^{\,(2)\rm s}(x)$ which contributes to the evolution
of the polarized valence quark distribution, thus completing the 
determination of the NNLO splitting functions for helicity-dependent
parton distributions of hadrons.

A {\sc Form} procedure of our alternative implementation of the scheme 
of Refs.\ \cite{g5LV,g5L,ZvN-F3}, as~well as {\sc Form} and {\sc Fortran} 
files of our results for $\Delta P_{\rm ns}^{\,(2)\rm s}$ can be 
obtained by downloading the source of this article from 
{\tt http://arxiv.org/} or from the authors upon request.

% Some-one to thank? 
% Specifically, no. Unspecifically, maybe, Eric, John, Werner? 

% ---------------------------------------------------------------------

\subsection*{Acknowledgments}

This work has been supported by
the {\it Deutsche Forschungsgemeinschaft} (DFG) through contract MO 
1801/1-1,
the {\it European Research Council}$\,$ (ERC) Advanced Grant no.~320651,
{\it HEPGAME} and the UK {\it Science \& Technology Facilities Council}%
$\,$(STFC) grant ST/L000431/1.

% ---------------------------------------------------------------------

{\small
\setlength{\baselineskip}{0.35cm}

}

\end{document}